\documentclass[prc,preprint,showpacs,preprintnumbers,amsmath,amssymb]{revtex4}

\usepackage[mathscr]{eucal}
\usepackage{graphicx}
\def\slr#1{\setbox0=\hbox{$#1$}           
   \dimen0=\wd0                                 
   \setbox1=\hbox{/} \dimen1=\wd1               
   \ifdim\dimen0>\dimen1                        
      \rlap{\hbox to \dimen0{\hfil/\hfil}}      
      #1                                        
   \else                                        
      \rlap{\hbox to \dimen1{\hfil$#1$\hfil}}   
      /                                         
   \fi}

\def\myvecint#1{\!\int\!\!\frac{d{#1}}{(2\pi)^3}\,}
\def\gev#1{ GeV${}^{#1}$}
\def\be{\begin{eqnarray}}
\def\ee{\end{eqnarray}}

\def\etal{\emph{et al.}}

\renewcommand{\theequation}%
    {\arabic{section}.\arabic{equation}}
\makeatletter \@addtoreset{equation}{section} \makeatother

\begin{document}

\preprint{BCCNT: 05/401/334}

\title{Relativistic Calculation of Pentaquark Widths}

\author{Hu Li}
\author{C. M. Shakin}
\email[email:]{cshakin@brooklyn.cuny.edu}

\affiliation{%
Department of Physics\\
Brooklyn College of the City University of New York\\
Brooklyn, New York 11210
}%

\author{Xiangdong Li}
\affiliation{%
Department of Computer System Technology\\
New York City College of Technology of the City University of New
York\\
Brooklyn, New York 11201 }%

\date{\today}

\begin{abstract}
We calculate the widths of the various pentaquarks in a
relativistic model in which the pentaquark is considered to be
composed of a scalar diquark and a spin $1/2$ triquark. We
consider both positive and negative parity for the pentaquark.
There is a single parameter in our model which we vary and which
describes the size of the pentaquark. We obtain quite small widths
for the decay $\Theta^+\rightarrow N+K^+$ and for
$\Theta_c^0\rightarrow P+D^{*-}$ consistent with the experimental
situation. For the sum of the decay widths for
$\bar{\Xi}^{--}\rightarrow \Xi^-+\pi^+$ and
$\bar{\Xi}^{--}\rightarrow \Sigma^-+K^-$ we find values of the
order of $4-8$ MeV for pentaquarks of the characteristic size
considered in this work. (The experimental situation with respect
to te observation of the $\bar{\Xi}^{--}$ is somewhat uncertain at
this time.) We also provide results for the decays $N^+\rightarrow
N+\pi$ and $N_s^+\rightarrow\Lambda^0 +K^+$. Our model of
confinement plays an important role in our analysis and makes it
possible to use Feynman diagrams to describe the decay of the
pentaquark.
\end{abstract}


\maketitle

\section{INTRODUCTION}

There has been a great deal of interest in the study of
pentaquarks and a large number of experiments have been carried
out [1-11]. The $\Theta^+(1540)$ which decays to a kaon and a
nucleon has been seen in several experiments. It has been
interpreted as a pentaquark with a $udud\bar{s}$ structure [12]. A
pentaquark $\Theta_c^0$ with the assumed structure $udud\bar{c}$
has also been observed recently. In the present work we will
present calculations of the widths of several pentaquarks in a
relativistic diquark-triquark model which includes a model of
confinement that we have introduced previously in our study of
meson and nucleon structure.

A number of reviews have appeared. In Refs.[13,14] the
experimental evidence for the pentaquark is reviewed. Theoretical
and experimental developments are reviewed in Refs.[15-18]. A
number of theoretical papers have also appeared [12,19-24]. We are
particularly interested in the work of Ref. [24] in which a
diquark-triquark model is introduced to describe the pentaquark.
We will make use of a variant of that model in the present work,
since that model lends itself to the analysis of the pentaquark
decay using Feynman diagrams. The diagram we consider is presented
in Fig.1. There the pentaquark is represented by the heavy line
with momentum $P$. The pentaquark is composed of a diquark of
momentum $-k+P_N$ and a triquark of momentum $P+k-P_N$. The
triquark emits a quark ($u$, $d$ or $s$) which combines with the
diquark to form a baryon of momentum $P_N$. The final-state meson
of momentum $P-P_N$ is emitted along with the quark at the
triquark vertex. (In the simplest model the triquark could be
considered as composed of the final-state meson coupled to the
exchanged quark, which ultimately forms part of the final-state
baryon. [See Fig. 1.])

\begin{figure}
\includegraphics[bb=0 355 519 623, angle=0, scale=0.7]{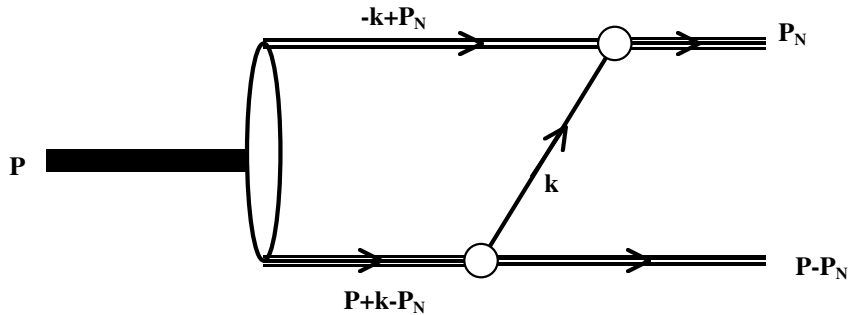}%
\caption{\label{f333.1}In this figure the heavy line denotes the
pentaquark and the line of momentum $-k+P_N$ denotes an
on-mass-shell diquark. The line of momentum $k$ represents the
quark and $P+k-P_N$ is the momentum of the triquark. In the final
state we have a baryon of momentum $P_N$ and a meson of momentum
$P-P_N$.}
\end{figure}

We have studied the structure of the nucleon in a quark-diquark
model in Ref. [25]. In that work we considered both scalar and
axialvector diquarks, however, in this work we will limit our
considerations to a nucleon composed of a quark and a scalar
diquark. The nucleon vertex is described in Ref. [25] for the case
in which the diquark is placed on mass shell. (We remark that in
Fig.1 only the quark of momentum $k$ and the triquark of momentum
$P+k-P_N$ will be off mass shell in our analysis.) The mass of the
scalar diquark was taken to be $400$ MeV in Ref.[25]. We use that
value here when the diquark does not have a strange quark present.
It is clear that we need to introduce a model of confinement for
the pentaquark to carry out our program. Our (covariant)
confinement model which we have used extensively in other works
[26-28] will be discussed in the next section.

The organization of our work is as follows. In Section II we will
describe our model of confinement and in Section III we will
discuss the widths of various pentaquarks of positive and negative
parity. In Section IV we will provide the results of our numerical
calculations and Section V contains some further discussion. The
Appendix contains a discussion of the normalization of the various
wave functions we have introduced.

\section{A Model of Confinement}

For the moment, let us consider the case of the $\Theta^+$
pentaquark with mass $1540$ MeV. Our pentaquark is composed of a
scalar diquark of mass $400$ MeV and a triquark of mass $800$ MeV.
Since the mass of the pentaquark under consideration is $1540$
MeV, the pentaquark would decay into its constituents in the
absence of a model of confinement. Similarly, the nucleon of mass
$939$ MeV could decay into the scalar diquark of mass $400$ MeV
and a quark whose mass we take to be $350$ MeV in this work.

In earlier work we have introduced a confining interaction which
served to prevent the decay of mesons or nucleons into their
constituents. Our covariant confinement model is described in a
series of our papers [26-28]. In that model we solve a linear
equation for a confining vertex function, $\Gamma$. This function
has the following property. Consider the decay $A\rightarrow B+C$,
in which the hadrons $A$ and $C$ are on mass shell. If we include
the confining vertex we find the amplitude has a zero when
particle $B$ goes on mass shell, so that the amplitude for $A$ to
decay into two on-mass-shell particles ($B$ and $C$) is zero. That
feature may be seen in Fig.2 which is taken from Ref. [29].

\begin{figure}
\includegraphics[bb=40 265 280 435, angle=0, scale=1]{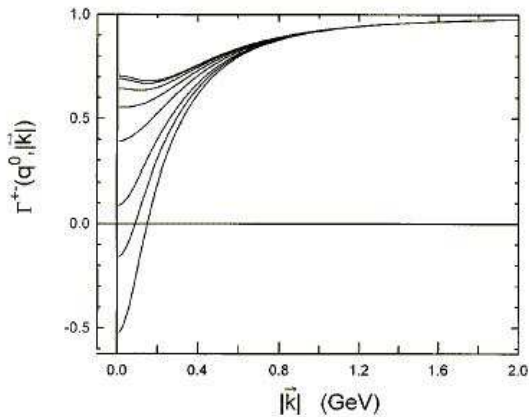}%
\caption{\label{f333.2}Values of
$\Gamma_S^{+-}(P^0,\mid\vec{k}\mid)$ defined in Ref. [29] are
shown. Starting with the uppermost curve and moving downward, the
values of $P^0$ are $0$, $0.10$, $0.20$, $0.30$, $0.40$, $0.50$,
$0.55$ and $0.60$ GeV. For the last two of these curves
$\Gamma_S^{+-}(P^0, k_{on})=0$. Here $k_{on}^2=(P_0^2/2)^2-m_q^2$,
with $m_q=0.260$ GeV.}
\end{figure}

For the decay $A\rightarrow B+C$ we may introduce a wave function
that may be expressed in terms of the momentum of the off-shell
particle $B$. If $B$ is a scalar, we have \be\label{e333.2.1}
\Psi_B(k)=\frac1{k^2-m_B^2}\Gamma_B(k)\,.\ee Note that the ratio
of $\Gamma_B(k)$ to $(k^2-m_B^2)$ is an ordinary function which
may often be well represented by a Gaussian function. (It is not
necessary to include an $i\epsilon$ in the denominator of Eq.
(\ref{e333.2.1}).)

In the case of the nucleon we may consider the decay into a quark
and a diquark. In Ref. [25] we considered both scalar and
axialvector diquarks, but for simplicity we will limit ourselves
to only the scalar diquark. The relevant wave function in this
case was given in Eq. (3.3) of Ref. [25]:
\be\Psi_S(P,k,s,t)=\widetilde\Psi_{(1)}(P,k)\frac{2m_q\Lambda^{(+)}(\vec
k)}{\sqrt{2E(\vec k)\left(E(\vec k)+m_q\right)}}u_N(P,s)\chi_t\ee
which we will simplify for the present work to read
\be\label{e333.2.3}\Psi_N(P_N,k,s,t)=\widetilde\Psi_N(P_N,k)\Lambda^{(+)}(\vec
k)u_N(P_N,s)\chi_t\,.\ee The function $\widetilde\Psi_N(P,k)$ is
represented in Fig. 5 of Ref. [25] by a dashed line. That function
is well approximated by a Gaussian function which we will record
at a latter point in our discussion. The factor of
$\Lambda^{(+)}(\vec k)=(\slr k_{on}+m_q)/2m_q$ arises from an
approximation made in Ref.[25]. (Here $k_{on}=(E(\vec k),\,\vec
k)$ with $E(\vec k)=[\vec k^{\,2}+m_q^2]^{1/2}$.) In Ref.[25] the
quark propagator was written as \be
-iS(k)=\frac{m_q}{E_q(k)}\left[\frac{\Lambda^{(+)}(\vec
k)}{k^0-E_q(\vec k)}-\frac{\Lambda^{(-)}(-\vec k)}{k^0+E_q(\vec
k)}\right]\,.\ee The second term was neglected in our formalism
when we studied the nucleon. (Thus we limited our analysis to
positive-energy quark spinors.) Since we wish to make use of the
nucleon wave function determined in Ref.[25], we will continue to
include the projection $\Lambda^{(+)}(\vec k)$ in our formalism.
Here $k$ is the quark momentum.

As stated earlier, the diquark of momentum $-k+P_N$ in Fig.
\ref{f333.1} will be placed on mass shell, so that $k^0=E_N(\vec
P_N)-E_D(\vec P_N-\vec k)$, where $E_D(\vec P_N-\vec k)=[(\vec
P_N-\vec k)^{\,2}+m_D^2]^{1/2}$ and $E_N(\vec P_N)=[\vec
P_N^{\,2}+m_N^2]^{1/2}$. That approximation is achieved by writing
\be \frac1{(P_N-k)^2-m_D^2+i\epsilon}\longrightarrow -2\pi
i\delta^{(+)}[(P_N-k)^2-m_D^2]\,,\ee as described in detail in
Ref.[30]. Note that \be
\delta^{(+)}[(P_N-k)^2-m_D^2]=\frac1{2E_D(\vec P_N-\vec
k)}\,\delta[P_N^0-k^0-E_D(\vec P_N-\vec k)]\,.\ee The
on-mass-shell specification used here arises when performing an
integral in the complex $k^0$ plane [30].

\section{Calculation of the Widths of Negative and Positive Parity Pentaquarks}

We consider the diagram shown in Fig. \ref{f333.1}. Recall that
the heavy line of momentum $P$ denotes the pentaquark. The line
carrying momentum $-k+P_N$ is the on-mass-shell scalar diquark and
the line with momentum $P+k-P_N$ is the triquark. The momentum $k$
is that of an up, down or strange quark, $P-P_N$ is the momentum
of the final-state baryon. The pentaquark, final-state nucleon and
final-state meson are all on mass shell. In our analysis the
diquark is also on mass shell, so that only the triquark and quark
propagate off mass shell, as noted earlier. (As stated in the last
section, the on-shell characterization of the diquark arises when
we complete the $k^0$ integral in the complex $k^0$ plane.)

We now make use of the formula [31] \be
d\Gamma=|\mathfrak{M}|^2\frac{d^3k_1}{(2\pi)^3}\,\frac{m_N}{E_N(\vec
k_1)}\,\frac{d^3k_2}{(2\pi)^3}\,\frac1{2E_K(\vec k_2)}\,
(2\pi)^4\delta^4(P-k_1+k_2),\ee where $\vec k_1$ and $\vec k_2$
are the momenta of the outgoing particles. We may put $\vec
k_1=\vec P_N$ and $\vec k_2=\vec P-\vec P_N=-\vec P_N$ for a
pentaquark at rest. (It is convenient to take $\vec P_N$ along the
$z$ axis when calculating the width.)

In writing our expression for $\Gamma$ we will represent the
product of the quark propagator and nucleon vertex function by the
nucleon wave function of Eq. (2.3). In a similar fashion we will
represent the product of the triquark propagator and the
pentaquark-triquark vertex function by a triquark wave function.
(In the vertex, the diquark is on mass shell.) We then have to
specify the vertex function of the triquark which describes the
decay into the quark of momentum $k$ and the final-state meson.
That scalar part of the vertex is usefully written as \be
\Gamma_T(k)=\Psi_T(k)(k^2-m_q^2)\,.\ee Thus, the wave functions
$\Psi_N(k)$, $\Psi_T(k)$ and $\Psi_\Theta(P_N-k)$ will appear in
our expression for $\Gamma_\Theta$ when we consider the decay of
the $\Theta^+$. Note that for the decay $\Theta^+\rightarrow
N+K^+$, we have \be \vec
P_N^2=\left(\frac{m_\Theta^2-m_N^2+m_K^2}{2m_\Theta}\right)^2-m_K^2\ee
which yields $\vec P_N^2=0.0722$ \gev2, or $|\vec P_N|=0.269$ GeV.

We find that the width is given by
\be\label{e333.3.4}\Gamma_\Theta&=&\frac12\myvecint {\vec
k}\frac1{2E_D(\vec k-\vec P_N)}\myvecint{\vec
k^{\,\prime}}\frac1{2E_D(\vec k^{\,\prime}-\vec
P_N)}\frac1{N_\Theta N_T N_N}\frac1{(2m_N)(2m_\Theta)}\\\nonumber
&\times&\Psi_N(\vec k)\,\Psi_N(\vec
k^{\,\prime})\,\Psi_\Theta(\vec P_N-\vec k)\,\Psi_\Theta(\vec
P_N-\vec k^{\,\prime})\,(k^2-m_q^2)\Psi_T(\vec
k)\,(k^{\prime2}-m_q^2)\Psi_T(\vec k^{\,\prime})\\\nonumber
&\times&\mbox{Tr}\left[\left(\slr k-\slr P_N+\slr
P+m_T\right)\left(\slr k_{on}+m_q\right)\left(\slr
P_N+m_N\right)\gamma^0\right.\\\nonumber &&\left. \left(\slr
k_{on}^\prime+m_q\right)\left(\slr k^\prime-\slr P_N+\slr
P+m_T\right)\gamma^0\left(\slr P+m_\Theta\right)\right]\rho\,,\ee
where $\rho$ is the phase space factor. We find for the decay of
the $\Theta^+$, \be \rho=\frac1{2\pi}\frac{m_N}{m_\Theta}|\vec
P_N|\\=0.0261 \,\;\mbox{GeV}\,,\ee since $|\vec P_N|\simeq0.269$
GeV.

The factor $\slr P_N+m_N$ and $\slr P+m_\Theta$ in the trace arise
from the relations \be\frac{\slr P_N+m_N}{2m_N}=\sum_{s_N}u_N(\vec
P_N,s_N)\bar u_N(\vec P_N,s_N)\ee and \be\frac{\slr
P+m_\Theta}{2m_\Theta}=\sum_s u_\Theta(\vec P,s)\bar u_\Theta(\vec
P,s)\,.\ee The factors $(1/N_\Theta)^{1/2}$, $(1/N_T)^{1/2}$ and
$(1/N_N)^{1/2}$ serve to normalize the wave functions. (See the
Appendix.) We determine that $N_N=0.316$, $N_T=0.0982$, and
calculate $N_\Theta$ for each choice of the pentaquark wave
function. We write, with $k=|\vec k|$, \be\Psi_N(\vec
k)=\frac1{\sqrt{N_N}}\left(y_0+\frac
A{w_0\sqrt{\dfrac\pi2}}\,\,e^{-\dfrac{2(k-k_c)^2}{w_0^2}}\right)\ee
and \be\Psi_T(\vec k)=\frac1{\sqrt{N_T}}\left(y_0+\frac
A{w_0\sqrt{\dfrac\pi2}}\,\,e^{-\dfrac{2(k-k_c)^2}{w_0^2}}\right)\,.\ee
We have determined $y_0$, $A$, $k_c$ and $w_0$ from a fit to the
wave function given in Fig. 5 of Ref.[25]. We find $y_0=-3.66$,
$k_c=-0.013$ GeV, $w_0=0.660$ GeV and $A=27.73$ GeV. For the
pentaquark we write \be\Psi_\Theta(\vec k-\vec
P_N)=\frac1{\sqrt{N_\Theta}}\frac
A{w_\Theta\sqrt{\dfrac\pi2}}\,\,e^{-\dfrac{2(\vec k-\vec
P_N)^2}{w_\Theta^2}}\,,\ee where $w_\Theta$ is a variable in our
analysis. (Note that $N_\Theta$ depends upon the choice of
$w_\Theta$.) Here $\vec k-\vec P_N$ is the relative momentum of
the diquark and triquark when $\vec P=0$.

In the case of a positive-parity pentaquark we assume that the
pentaquark decays to a positive-parity diquark and a
positive-parity triquark. As before, the triquark and diquark have
zero relative angular momentum. In this case we need to insert
factors of $i\gamma_5$ at the triquark-meson vertex where the
quark of momentum $k$ is emitted. In addition, the calculation of
the normalization factor $N_T$ is modified. (See the Appendix.) In
Eq. (3.4), the trace becomes \be
\mbox{Trace}&=&\mbox{Tr}\left[(\slr k-\slr P_N+\slr
P+m_T)i\gamma_5(\slr k_{on}+m_q)(\slr
P_N+m_N)\gamma^0\right.\\\nonumber &&\left. (\slr
k_{on}^\prime+m_q)i\gamma_5(\slr k^\prime-\slr P_N+\slr
P+m_T)\gamma^0(\slr P+m_\Theta)\right]\,.\ee We may define
$\widetilde{ \slr k}=\gamma^0\slr k\gamma^0$, etc. Thus \be
\mbox{Trace}&=&\mbox{Tr}\left[(\slr k-\slr P_N+\slr
P+m_T)\gamma_5(\slr k_{on}+m_q)(\slr P_N+m_N)(\widetilde{\slr
k}_{on}^{\,\prime}+m_q)\right.\\\nonumber &&\left.
\gamma_5(\widetilde{\slr k}^{\,\prime}-\widetilde{\slr
P}_N+\widetilde{\slr P}+m_T)(\slr
P+m_\Theta)\right] \\
&=& \mbox{Tr}\left[(\slr k-\slr P_N+\slr P+m_T)(-\slr
k_{on}+m_q)(-\slr P_N+m_N)(-\widetilde{\slr
k}_{on}^{\,\prime}+m_q)\right.\\\nonumber &&\left.
(\widetilde{\slr k}^{\,\prime}-\widetilde{\slr
P}_N+\widetilde{\slr P}+m_T)(\slr P+m_\Theta)\right]\,.\ee

In this case we find $N_N=0.316$ and $N_T=0.0201$, where only the
second value has changed relative to the values given above.

\section{Numerical Results}

In this section we present the results of our numerical
calculations. In Fig. 3 we exhibit the width for the decay
$\Theta^+\rightarrow N+K^+$ in the case that the pentaquark has
positive parity. The result for a negative-parity pentaquark is
shown in Fig. 4. The parameters used are given in the caption of
Fig. 3. The small value for the diquark mass used here corresponds
to the analysis made in Ref. [29]. In the case of the negative
parity pentaquark (Fig. 4) very small widths may be obtained for
$w$ in the range $0.75$ GeV $<w<0.80$ GeV. For the case of
positive parity, quite small widths are found in the range $0.70$
GeV $<w< 0.90$ GeV. [See Fig. 3.]

\begin{figure}
\includegraphics[bb=0 0 280 235, angle=0, scale=1]{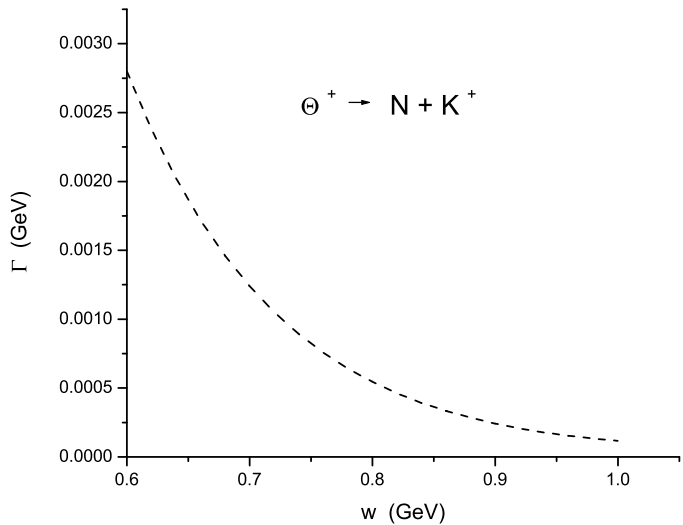}%
\caption{\label{f333.3}The width for the decay
$\Theta^+\rightarrow N+K^+$ for a positive-parity pentaquark as
calculated in our model is shown. Here $m_{\Theta}=1.540$ GeV,
$m_T=0.800$ GeV, $m_D=0.400$ GeV, $m_N =0.939$ GeV and $m_K
=0.495$ GeV.}
\end{figure}

\begin{figure}
\includegraphics[bb=0 0 280 235, angle=0, scale=1]{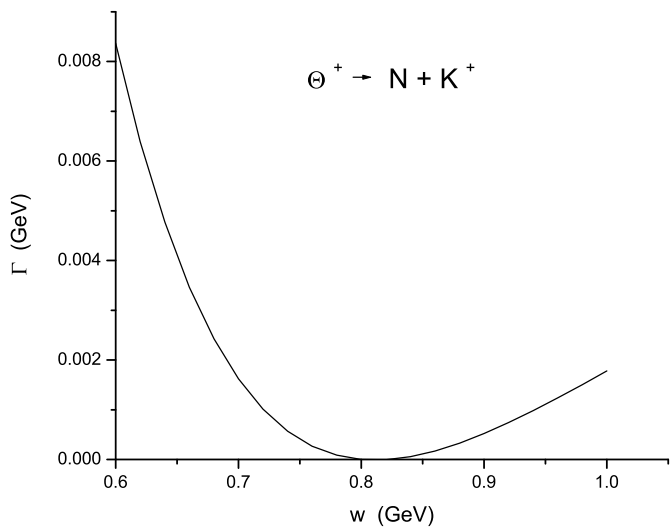}%
\caption{\label{f333.4}The width of the negative parity pentaquark
is shown as a function of the parameter $w$. (The quite small
values in the vicinity of the minimum have rather large
uncertainties because of the limitation of the number of points
used in our five-dimensional integral which determines the width.)
The parameters used are given in the caption to Fig. 3.}
\end{figure}
\begin{figure}
\includegraphics[bb=0 0 280 235, angle=0, scale=1]{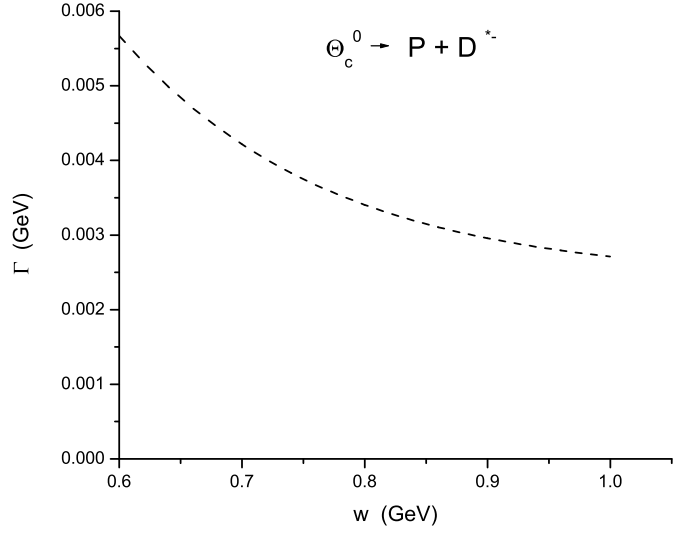}%
\caption{\label{f333.5}The calculated width of the positive-parity
charmed pentaquark is shown for the decay $\Theta_c^0\rightarrow
P+D^{*-}$. Here. $m_{\Theta_c^0}$=3.099 GeV, $m_P$=0.939 GeV,
$m_D=0.400$ GeV, $m_T=2.00$ GeV, $m_q=0.350$ GeV and
$m_{D^{*-}}=2.007$ GeV.}
\end{figure}

\begin{figure}
\includegraphics[bb=0 0 280 235, angle=0, scale=1]{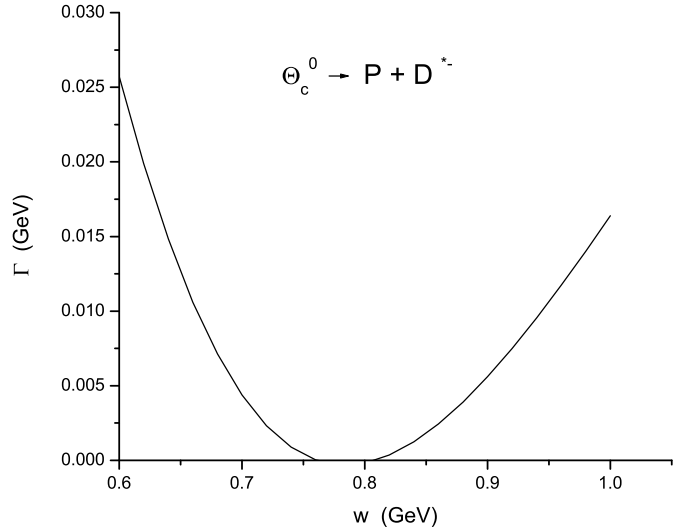}%
\caption{\label{f333.6}The calculated width of the negative-parity
charmed pentaquark is shown. The calculation is made for the
parameters given in the caption of Fig. 5.}
\end{figure}

\begin{figure}
\includegraphics[bb=0 0 280 235, angle=0, scale=1]{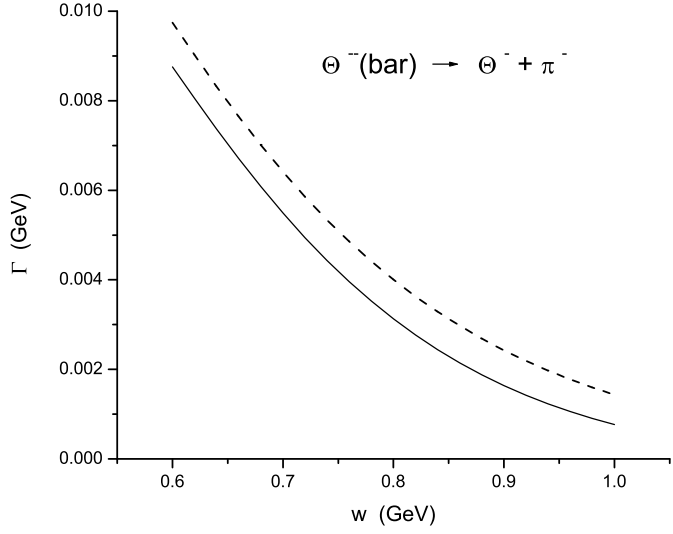}%
\caption{\label{f333.7}The calculated width for the decay
$\bar{\Xi}^{--} \rightarrow\Xi^-+\pi^-$ is shown for the case that
the $\bar{\Xi}^{--}$ has positive parity (dashed line) and for
negative parity (solid line). The parameters used are
$m_{\bar{\Xi}}$=1.862 GeV, $m_T$=0.800 MeV, $m_D=0.700$ GeV,
$m_s=0.450$ GeV, $m_{\Xi}=1.321$ GeV and $m_\pi=0.139$ GeV.}
\end{figure}

\begin{figure}
\includegraphics[bb=0 0 280 235, angle=0, scale=1]{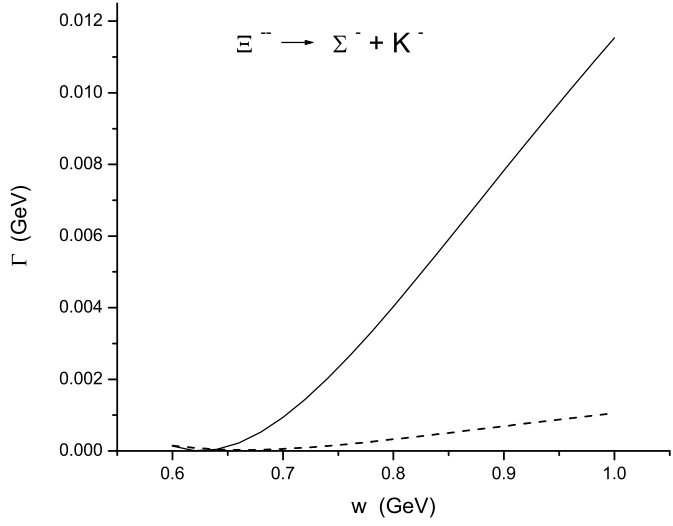}%
\caption{\label{f333.8}The calculated width for the decay
$\bar{\Xi}^{--}\rightarrow\Sigma^-+K^-$ is shown for the case that
the $\bar{\Xi}^{--}$ has positive parity (dashed line) and for
negative parity (solid line). The parameters used for
$m_{\bar{\Xi}}$, $m_T$, $m_D$, and $m_q$ are given in the caption
to Fig. 7. Here $m_{\Sigma^-}=1.197$ GeV and $m_K = 0.495$ GeV.}
\end{figure}

\begin{figure}
\includegraphics[bb=0 0 280 235, angle=0, scale=1]{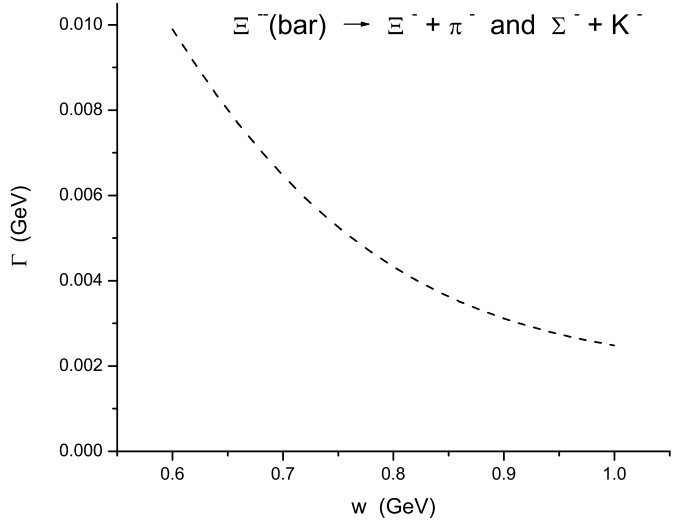}%
\caption{\label{f333.9}The calculated widths for the decays
$\bar{\Xi}^{--}\rightarrow\Xi^-+\pi^-$ and $\Sigma^-+K^-$ are
summed and shown in the figure. Here the $\bar{\Xi}^{--}$ is
assumed to have positive parity. The parameters used are given in
the captions of Figs.7 and 8.}
\end{figure}

\begin{figure}
\includegraphics[bb=0 0 280 235, angle=0, scale=1]{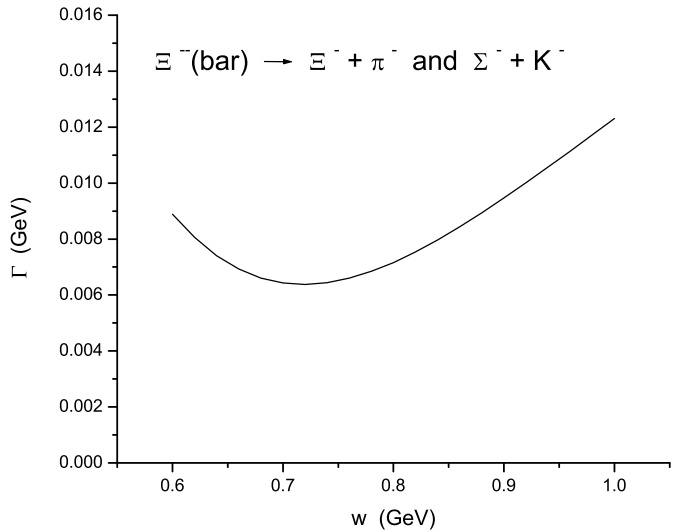}%
\caption{\label{f333.10}Same caption as Fig. 9 except that the
$\bar{\Xi}^{--}$ is taken to have positive parity.}
\end{figure}

\begin{figure}
\includegraphics[bb=0 0 280 235, angle=0, scale=1]{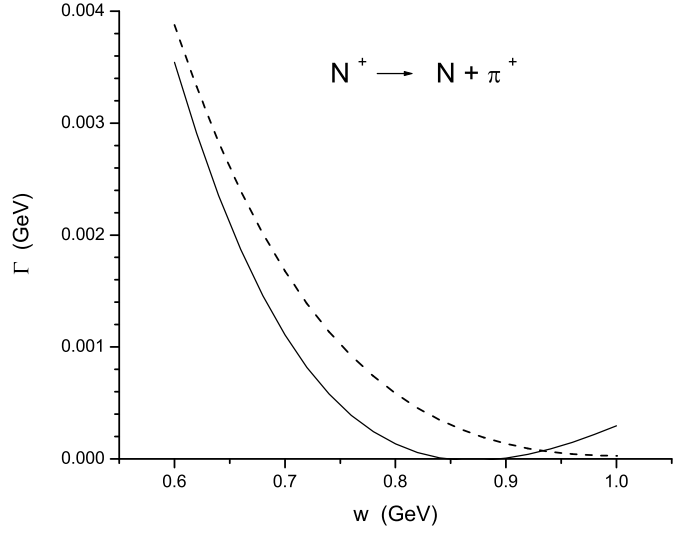}%
\caption{\label{f333.11}The values obtained for the decays
$N^+\rightarrow N+\pi^+$ are shown for a positive-parity $N^+$
(dashed line) and for negative parity (solid line). The parameters
used were $m_N=0.939$ GeV, $m_\pi=0.139$ GeV, $m_{N^+}=1.440$ GeV,
$m_T=0.800$ GeV and $m_d=0.350$ GeV.}
\end{figure}

\begin{figure}
\includegraphics[bb=0 0 280 235, angle=0, scale=1]{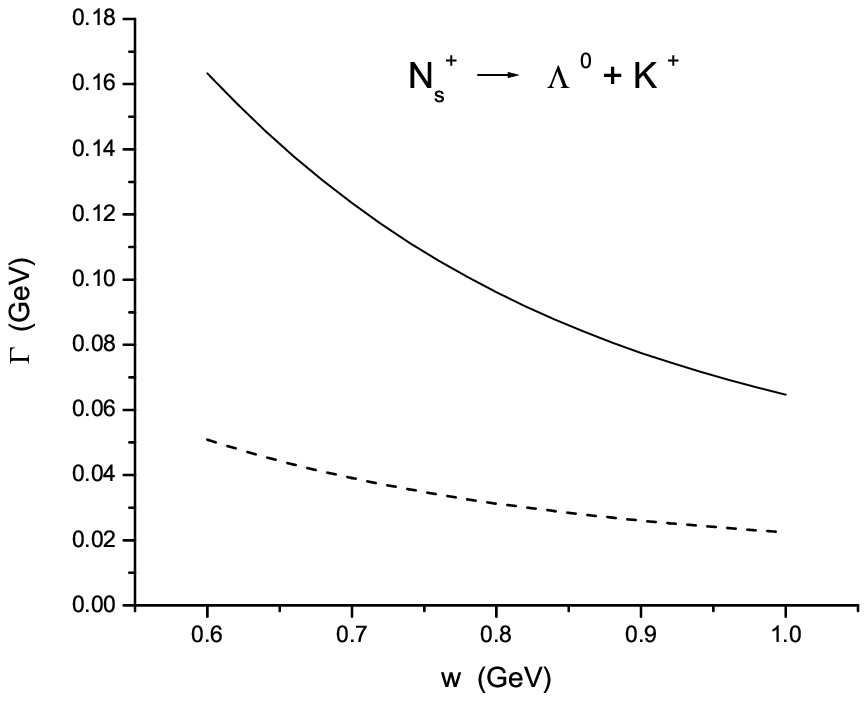}%
\caption{\label{f333.12}The calculated width for the decays
$N_s^+\rightarrow\Lambda^0+K^+$ are shown for a positive parity
$N_s^+$ (dashed line) and for negative parity (solid line). The
parameters used were $m_{\Lambda^0}=1.116$ GeV, $m_{N_s}=1.700$
GeV, $m_D=0.400$ GeV, $m_s=0.450$ GeV, $m_K=0.495$ GeV and
$m_T=1.00$ GeV.}
\end{figure}

Results for the decay $\Theta_c^0\rightarrow P+D^{*-}$ are given
in Figs. 5 and 6. Again, we find very small widths for the
negative-parity pentaquark in the range $0.75$ GeV$<w<0.80$ GeV
with widths of about $3\rightarrow 4$ MeV for the positive-parity
$\Theta_c^0$ in the corresponding range. (The parameters used are
given in the figure captions.)

As stated earlier, the observation of the $\bar{\Xi}^{--}$
pentaquark is a matter of some controversy. Its observation was
reported in Ref. [3] with a width less than $18$ MeV, however,
that state was not seen in the work reported in Refs. [5] and
[11]. Our results calculated for the decay
$\bar{\Xi}^{--}\rightarrow \Xi^-+\pi^-$ are shown in Fig. 7, while
our results for the decay $\bar{\Xi}^{--}\rightarrow \Sigma^-+K^-$
are given in Fig. 8. In the case of a negative parity
$\bar{\Xi}^{--}$ the summed results for the two decay channels are
given in Fig.\,9 (negative-parity pentaquark) and Fig.\,10
(positive-parity pentaquark). The calculated widths are of the
order of $4$-$8$ MeV in the region $0.75$ GeV$<w<0.80$ GeV. The
$\bar{\Xi}^{--}$ is not seen in some experiments that involved a
search for this pentaquark [5, 11].

In Fig. 11 we present our results for the decay $N^+\rightarrow
N+\pi^+$. Here we use the mass suggested in Ref. [12] for the
$N^+$. In this case very small widths are calculated. Much larger
widths for the $N^+$ decay to $\Lambda^0+K^+$ are found and the
calculated values are exhibited in Fig. 12. Since the $N^+$ and
$N_s^+$ have the quantum numbers of the nucleon, they are expected
to mix with neighboring states and be hard to distinguish from
other nucleon resonances. The mixing of such states is discussed
in Ref. [32].

\section{discussion}

It is of interest to note that the same computer code can yield
different values for the pentaquark widths depending upon the mass
values inserted for the pentaquark, final-state baryon and meson,
and for the diquark and triquark. We do recognize that our model
is limited since we have not taken into account the full symmetry
of the wave function. That is, we do not consider the identity of
the quarks in the diquark and the triquark. It is possible that a
more complex calculation made in the future may overcome this
limitation of the model. On the other hand, our results are quite
suggestive, since very small widths are obtained for the
$\Theta^+$ and $\Theta_c^0$ in accordance with the experimental
situation. It is clear that further experimental studies and
theoretical analysis is desirable.

\appendix
  \renewcommand{\theequation}{A\arabic{equation}}
  \setcounter{equation}{0}  
  \section{Normalization parameter for the negative-parity
  and positive-parity pentaquark states}  

The normalization of the final-state baryon wave function is
calculated by requiring that that the baryon contain a single
quark in addition to the scalar diquark. In that calculation the
diquark appears on mass shell. We find in the case of a nucleon in
the final state, the normalization parameter is \be N_N =
\frac{1}{2\pi^2}\int_0^{k_{max}}\frac{k^2dk}{E_D(\vec{k})}(kk^0)(k^0+m_q)\mid\Psi_N(\vec{k})\mid^2,\ee
where $E_D(\vec{k})=(\vec{k}^2+m_D^2)^{1/2}$, with $m_D$ being the
mass of the diquark and $k^0=m_N-E_D(\vec{k})$. Here we take
$k_{max}=0.7$ GeV. In Eq. (A1) \be
\Psi_N(\vec{k})=\left(y_0+\frac{A}{w_0\sqrt{\frac{\pi}{2}}}\exp^{-\frac{2(k-k_c)^2}{w_0^2}}\right).
\ee [See Eq. (3.9).] Values of $y_0, w_0, k_c$ and $A$ are given
after Eq. (3.10).

The calculation of the normalization factor for the pentaquark
yields a similar result. For example \be
N_\Theta=\frac{1}{2\pi^2}\int_0^{k_{max}}\frac{k^2dk}{E_D(\vec{k})}k^0k(k^0+m_T)\mid\Psi_\Theta(\vec{k})\mid^2,
\ee with $k^0=m_\Theta-E_D(\vec{k})$. Here
$E_D(\vec{k})=(\vec{k}^2+m_D^2)^{1/2}$, with $m_D$ being the mass
of the diquark. For the pentaquark, we use the wave function \be
\Psi_\Theta(\vec{k})=\frac{A}{w\sqrt{\frac{\pi}{2}}}\exp^{-\frac{2(\vec{k}-\vec{P}_N)^2}{w^2}}
\ee in Eq. (A3). [See Eq. (3.11).] Here $w$ is a variable of our
model. Results for various values of $w$ are given in the figures
for different pentaquarks.

The triquark normalization factor is calculated by requiring that
the triquark contain a single quark in addition to the final-state
meson which is placed on mass shell in the calculation. We find
\be
N_T=\frac{1}{4\pi^2}\int_0^{k_{max}}k^2dk(k_0^2+\vec{k}^2+2k_0m_q+m_q^2)\mid\tilde{\Psi}_T(\vec{k})\mid^2,\ee
where $k_0=m_T-E_K(\vec{k})$. Here $E_K=(\vec{k}^2+m_K^2)^{1/2}$,
where $m_K$ is the mass of the final-state meson, and $m_q$ is the
mass of the exchanged quark. [See Fig. 1.] Note that, in this
case, \be
\Psi_T(\vec{k})=y_0+\frac{A}{w_0\sqrt{\frac{\pi}{2}}}\exp^{-\frac{2(k-k_c)^2}{w_0^2}},
\ee with the parameters $y_0$, $w_0$, $k_c$ and $A$ given after
Eq. (3.10).

If we consider positive parity pentaquark states, we do not have
to change the baryon normalization parameter which was given in
Eq. (A1).

The normalization parameter of the pentaquark wave function is
still of the form given in Eq. (A2), while the triquark
normalization factor is now \be
N_T=\frac{1}{4\pi^2}\int_0^{k_{max}}k^2dk(k_0^2+\vec{k}^2-2k_0m_q+m_q^2)\mid\tilde{\Psi}_T(\vec{k})\mid^2.\ee
Comparison to Eq. (A5) shows that the sign of the term $2k_0m_q$
in Eq. (A7) has changed relative to the sign in Eq. (A5).



\end{document}